\documentclass[10pt,conference]{IEEEtran}
\IEEEoverridecommandlockouts
\usepackage{cite}
\usepackage{amsmath,amssymb,amsfonts,mathrsfs,subfigure,tcolorbox,framed}
\usepackage[hyphens]{url}
\usepackage{algorithmic}
\usepackage[hidelinks]{hyperref}
\usepackage{graphicx}
\usepackage{textcomp}
\usepackage{xcolor}
\usepackage{balance}
\usepackage{adjustbox,colortbl}
\def\BibTeX{{\rm B\kern-.05em{\sc i\kern-.025em b}\kern-.08em
    T\kern-.1667em\lower.7ex\hbox{E}\kern-.125emX}}

\begin{document}

\title{An Empirical Study on Deployment Faults of Deep Learning Based Mobile Applications}

\author{\IEEEauthorblockN{Zhenpeng Chen\IEEEauthorrefmark{1},
Huihan Yao\IEEEauthorrefmark{1},
Yiling Lou\IEEEauthorrefmark{1}, 
Yanbin Cao\IEEEauthorrefmark{1}\IEEEauthorrefmark{2},
Yuanqiang Liu\IEEEauthorrefmark{1}, 
Haoyu Wang\IEEEauthorrefmark{3}, and
Xuanzhe Liu\IEEEauthorrefmark{1}}
\IEEEauthorblockA{\IEEEauthorrefmark{1}Key Lab of High Confidence Software Technologies (Peking University), Ministry of Education, Beijing, China}
\IEEEauthorblockA{\IEEEauthorrefmark{2}Peking University Information Technology Institute (Tianjin Binhai), Tianjin, China}
\IEEEauthorblockA{\IEEEauthorrefmark{3}Beijing University of Posts and Telecommunications, Beijing, China}
\IEEEauthorblockA{\{czp, yaohuihan, yiling.lou, caoyanbin, yuanqiangliu\}@pku.edu.cn, haoyuwang@bupt.edu.cn, xzl@pku.edu.cn}% <-this % stops an unwanted space
\thanks{Corresponding author: Xuanzhe Liu (xzl@pku.edu.cn).}}

\newcommand{\para}[1]{\smallskip\noindent{\bf {#1}. }}
\newcommand{\czp}[1]{{\color{red}{#1}}}
\newcommand{\yiling}[1]{{\color{blue}{#1}}}
\newcommand{\xzl}[1]{{\color{purple}{#1}}}

\maketitle

\begin{abstract}
Deep learning (DL) is moving its step into a growing number of mobile software applications. These software applications, named as DL based mobile applications (abbreviated as \emph{mobile DL apps}) integrate DL models trained using large-scale data with DL programs. A DL program encodes the structure of a desirable DL model and the process by which the model is trained using training data. Due to the increasing dependency of current mobile apps on DL, software engineering (SE) for mobile DL apps has become important. However, existing efforts in SE research community mainly focus on the development of DL models and extensively analyze faults in DL programs.
In contrast, faults related to the deployment of DL models on mobile devices (named as \emph{deployment faults of mobile DL apps}) have not been well studied. Since mobile DL apps have been used by billions of end users daily for various purposes including for safety-critical scenarios, characterizing their deployment faults is of enormous importance. To fill in the knowledge gap, this paper presents the first comprehensive study to date on the deployment faults of mobile DL apps. We identify 304 real deployment faults from Stack Overflow and GitHub, two commonly used data sources for studying software faults. Based on the identified faults, we construct a fine-granularity taxonomy consisting of 23 categories regarding to fault symptoms and distill common fix strategies for different fault symptoms. Furthermore, we suggest actionable implications and research avenues that can potentially facilitate the deployment of DL models on mobile devices.

\end{abstract}


\section{Introduction}\label{intro}
In recent years, deep learning (DL) has emerged as one of the most popular and promising techniques and has been widely adopted in various applications~\cite{wwwMaXZTL19,wwwChenSHLML19,ZhouLKGZ0Z020,sigsoftChenCLML19,chentosem}. Mobile devices are undoubtedly among the most important platforms for running DL based software applications~\cite{mobileai1,mobileai2,mobileai3}. These software applications, namely DL based mobile applications (in short as \emph{mobile DL apps}), integrate DL capabilities to add a wide range of features including object detection~\cite{wang2018pelee}, image processing~\cite{mobicomXuZLLL18}, natural language processing~\cite{imwutXuQMHL18}, speech recognition~\cite{mustafa2019comparative}, etc. To achieve this goal, developers train DL models using large-scale data (i.e., \emph{development of DL models}), and then deploy the obtained DL models on mobile devices for real usage (i.e., \emph{deployment of DL models}).
% Due to the rapidly growing number of mobile DL apps~\cite{wwwXuLLLLL19}, software engineering (SE) for them has become increasingly important.
% , activity recognition~\cite{radu2016towards}

In fact, \emph{development of DL models} is a general process for different types of DL based applications~\cite{CHENDLDEPLOY} and its challenges have been well studied in the software engineering (SE) research community~\cite{eseHanSWDX20,issredeep19,isstaZhangCCXZ18,sigsoftIslamNPR19,corrabstaxonomy,icse20DLREPAIR}.
% Due to the rapidly growing number of mobile DL apps~\cite{wwwXuLLLLL19}, software engineering (SE) for them has become increasingly important. 
% including \emph{development} and \emph{deployment}~\cite{GuoCXMHLLZL19,CHENDLDEPLOY}. The development process aims to obtain DL models trained using large-scale data with DL programs, while the deployment process refers to deploying the obtained DL models on mobile devices. Many efforts in SE research community have been devoted to characterizing the challenges in the two processes. 
% Specifically, 
% Many efforts in SE research community have been devoted to characterizing challenges in the process. 
In particular, researchers~\cite{isstaZhangCCXZ18,sigsoftIslamNPR19,corrabstaxonomy,icse20DLREPAIR} have extensively analyzed faults in the DL programs written based on DL frameworks (e.g., TensorFlow (TF)~\cite{osdiAbadiBCCDDDGIIK16} and Keras~\cite{keraslink}), which encode the structure of desirable DL models and the process by which the models are trained using the training data. 
% In particular, researchers have extensively analyzed faults in the DL programs that encode the structure of desirable DL models and the process by which the models are trained using the training data. Since developers implement DL programs relying on DL frameworks such as TensorFlow (TF)~\cite{osdiAbadiBCCDDDGIIK16} and Keras~\cite{keraslink}, various studies~\cite{isstaZhangCCXZ18,sigsoftIslamNPR19,corrabstaxonomy,icse20DLREPAIR} focus on characterizing faults in DL programs written based on these frameworks and analyzing fix patterns of these faults.
% ,  PyTorch~\cite{pytorchlink}, Theano~\cite{RfouAAa16}, and Caffe~\cite{JiaSDKLGGD14},

Recently, the rapid growth  of mobile DL apps~\cite{wwwXuLLLLL19} has posed urgent challenges to the \emph{deployment of DL models}, i.e., deploying DL models on mobile devices. For example, computation-intensive DL models can be executed efficiently on PC/server platforms, but they cannot be directly deployed and executed on mobile devices with limited computing power~\cite{GuoCXMHLLZL19}. Although major vendors have rolled out specific DL frameworks such as TF Lite~\cite{tflitelink} and Core ML~\cite{coremllink} to facilitate this deployment process, various specific faults are still emerging in this process and frequently asked on Stack Overflow (SO), one of the most popular Q\&A forums for developers~\cite{CHENDLDEPLOY}. Moreover, previous work~\cite{CHENDLDEPLOY} has demonstrated that relevant questions are increasing rapidly on SO and more difficult to resolve than those related to other aspects of DL based applications. In addition, mobile DL apps are not only used by billions of end users for their daily activities (e.g., speech-to-text and photo beauty)~\cite{wwwXuLLLLL19,icdcsWangCYSBZ18}, but also reported to be increasingly adopted in various safety-critical scenarios (e.g., driver assistance~\cite{imwutChenS19} and autonomous vehicles~\cite{intelltran}). Therefore, the emerging faults related to the deployment of DL models on mobile devices (named as \emph{deployment faults of mobile DL apps}) should be carefully addressed. Unfortunately, the characteristics of these faults have not been well understood. 

% Recently, due to the rapidly growing number of mobile DL apps~\cite{wwwXuLLLLL19}, SE researchers~\cite{GuoCXMHLLZL19,CHENDLDEPLOY} begin to pay attention to the specific challenges in the \emph{deployment of DL models}, i.e., deploying DL models on mobile devices. Since computation-intensive DL models trained PC/server platforms cannot be directly deployed on mobile devices with limited computing power, major vendors have rolled out specific DL frameworks such as TF Lite~\cite{tflitelink} and Core ML~\cite{coremllink} to facilitate such a deployment process. Therefore, to characterize the challenges in deploying DL models on mobile devices, researchers focus on the usage of these frameworks. For example, Chen \textit{et al.}~\cite{CHENDLDEPLOY} summarized the challenges that developers face in the deployment process through manual inspection of posts related on TF Lite and Core ML on Stack Overflow (SO). They demonstrate that deploying DL models on mobile devices poses some specific programming challenges to developers such as converting DL models to the formats expected by mobile devices; these challenges frequently lead to faults and are asked on SO. Unfortunately, such faults related to deploying DL models on mobile devices (named as \emph{deployment faults of mobile DL apps}) have not been well studied. \xzl {(Here, you should articulate the importance of deployment faults. Why they matter? How difficult to identify deployment faults?)}

To fill in the knowledge gap, this paper presents the first comprehensive study on analyzing \emph{symptoms} and \emph{fix strategies} of deployment faults of mobile DL apps. Given the surging popularity of mobile DL apps, this study is of enormous importance. It can help in understanding what are the common deployment faults of mobile DL apps and how these faults are resolved in practice, so as to provide a high-level categorization that can serve as a guide for developers to resolve common faults and for researchers to develop tools for detecting and fixing deployment faults of the increasing mobile DL apps.
% provide guidance for preventing, detecting, localizing, and fixing of the deployment faults associated with the increasing mobile DL apps.

We focus our study on the faults that occur during the usage of two representative frameworks specifically designed for deploying DL models on mobile devices, i.e., TF Lite~\cite{tflitelink} and Core ML~\cite{coremllink}, both of which are widely used in industry practice and well adopted in related studies~\cite{GuoCXMHLLZL19,CHENDLDEPLOY}. Specifically, we collect a dataset of 304 deployment faults related to their usage from SO and GitHub, two commonly used data sources for studying software faults~\cite{isstaZhangCCXZ18,sigsoftIslamNPR19,corrabstaxonomy,icse20DLREPAIR,kbseFrancoGR17}. 
% Based on the dataset, we focus our study on the following questions that we believe could provide insights for developers and researchers.

By manual analysis, we qualitatively extract the symptom of each identified fault and construct a hierarchical taxonomy containing 23 symptom categories, indicating the diversity of deployment faults of mobile DL apps. Additionally, we distill common fix strategies for each symptom category, providing insights about deployment fault resolution of mobile DL apps. Based on our results, we discuss new directions for future research. Furthermore, we offer the scripts and the data used in this study~\cite{sumat} as an additional contribution to the research community for other researchers to replicate and build upon.

\section{Background and Research Questions}\label{back}
\begin{figure}
\includegraphics[width=1.0\columnwidth]{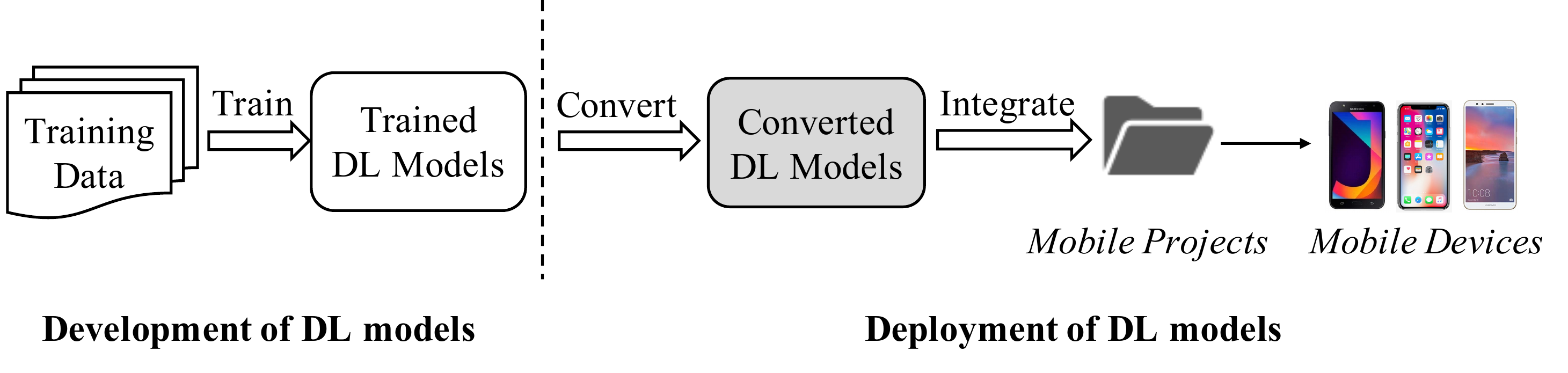}
\caption{Development and deployment of DL models}\label{fig:devordep}
\end{figure}

We start by introducing current practice of the development of DL models and the deployment of DL models on mobile devices. Fig.~\ref{fig:devordep} distinguishes the two processes.
% We start by introducing the current practice of the development and deployment processes of DL based software (in short as \emph{DL software}). Fig.~\ref{fig:devordep} distinguished the two processes.

\textbf{Development of DL models.} 
% To integrate DL capabilities into software applications, developers rely on state-of-the-art DL frameworks such as TF and Keras.
Development of DL models is a general process for different types of DL based software applications~\cite{CHENDLDEPLOY}. First, developers construct structures of desirable DL models and specify run-time configuration (e.g., hyper-parameters) with DL programs written based on state-of-the-art DL frameworks such as TF and Keras. A DL model consists of multiple layers to convert input to output, with each layer containing a set of neurons that accept input from neurons in the preceding layer, apply activation function to the input, and pass the resulting output to the neurons in the succeeding layer via a set of weighted edges. The layer used as an entry point into the DL model is called the input layer, while the layer that produces the end result is called the output layer. The input and output layers wrap the input and output tensors (multi-dimensional arrays of numerical values), respectively. Then, developers use large-scale data to train the DL models, during which the weights of edges in the models are adjusted and set to values that minimize the difference between model output and expected output.
Finally, developers evaluate the performance (e.g., accuracy) of the obtained DL models using testing data. Due to space limit, we present only the model training phase in Fig.~\ref{fig:devordep}.

\textbf{Deployment of DL models.} DL models, which are demonstrated to meet the performance requirements, are ready to be deployed on mobile devices for real usage. The deployment process mainly focuses on platform adaptations. 
Due to the limited computing power, memory size, and energy capacity of mobile devices, models trained on PC/server platforms cannot be directly deployed on them. To tackle this problem, some lightweight frameworks, such as TF Lite for Android and Core ML for iOS, are specifically designed for converting trained DL models to the formats supported by mobile devices. Specifically, Core ML provides Python APIs for this task, while TF Lite provides both CLIs and Python APIs. It is a common practice in the conversion stage to perform model quantization to reduce precision representations of the weights of edges in trained DL models, in order to reduce memory cost and computing overhead. For example, Core ML supports converting the weights from 32 bits to 16/8/4 bits. Then, developers can integrate the converted models into mobile projects with the help of TF Lite and Core ML. For instance, TF Lite provides APIs of various programming languages, such as Java, C++, and Python, to support the integration. Finally, the integrated projects can run on mobile devices and make inference based on input data. 
% Swift, Objective-C,

\textbf{Scope and research questions.} We focus our analysis on the process of deploying DL models to mobile devices. Any faults related to this process are within our scope. However, faults that occur during the development of DL models are not considered in this study. Specifically, we aim to address two research questions that are concerned with deployment faults of mobile DL apps:

\textbf{RQ1 (Symptoms):} \emph{What are the frequent fault symptoms?} 

% Understanding the kind and frequency of fault symptoms is the first step to characterize the deployment faults of mobile DL software, which is critical for triaging those faults and assessing their impacts. In addition, 
% As described before, the deployment process of mobile DL software involves several phases. Certain phases may be more prone to faults and are thus more important than others for fault identification an repair. This information further allows 

\textbf{RQ2 (Fix strategies):} \emph{What are the common fix strategies for different fault symptoms?}
\section{Methodology}\label{method}
\begin{figure}[h]
\centerline{\includegraphics[width=0.85\columnwidth]{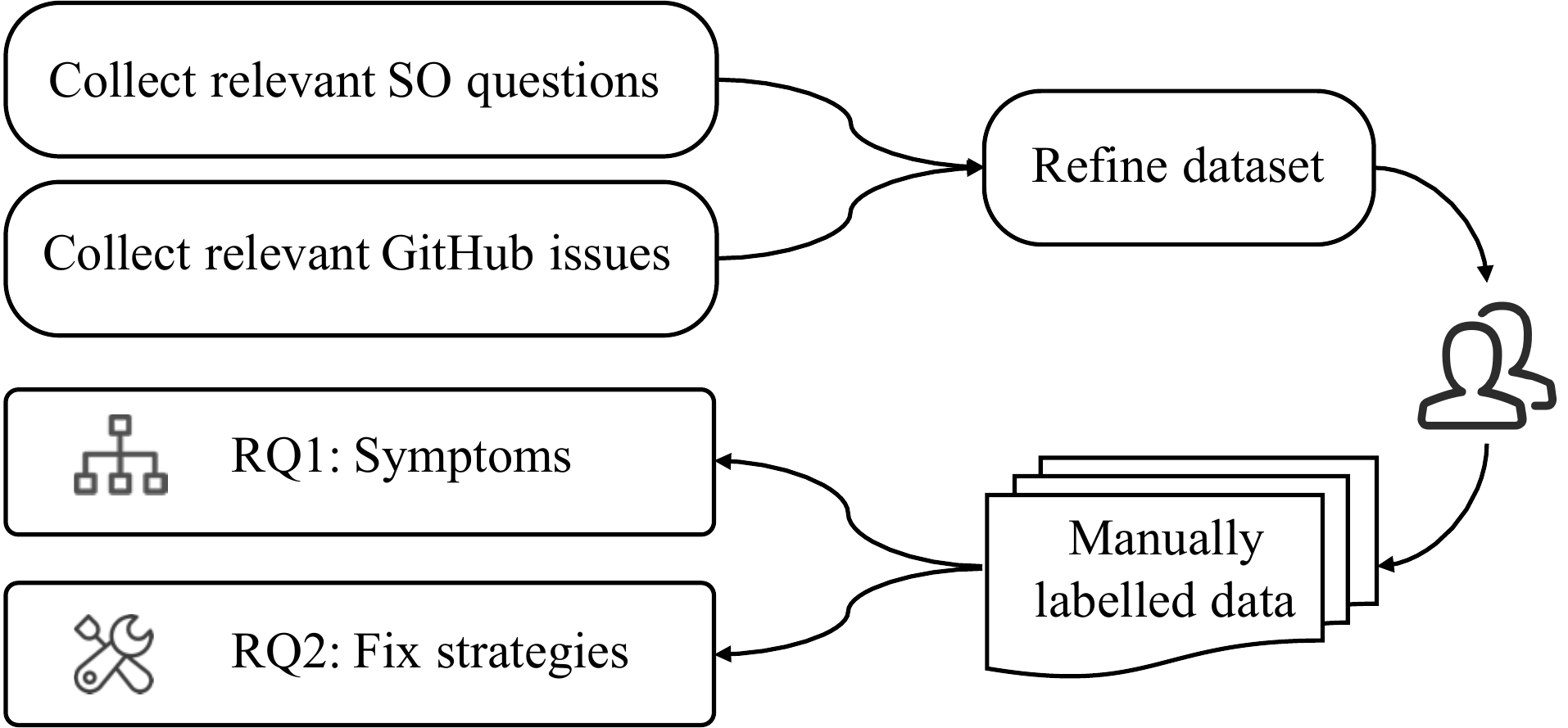}}
\caption{Overview of the methodology.}\label{fig:method}
\end{figure}

To characterize the deployment faults of mobile DL apps, we analyze the relevant questions posted on SO and the relevant issues posted on GitHub. 
% SO and GitHub are representative data sources in SE research community and have been widely adopted to study faults in different types of software systems in previous studies~\cite{isstaZhangCCXZ18,sigsoftIslamNPR19,corrabstaxonomy,icse20DLREPAIR,kbseFrancoGR17}. 
We illustrate the overview of the methodology of our study in Fig.~\ref{fig:method}.

\subsection{Data Collection}
% In line with previous work~\cite{isstaZhangCCXZ18,sigsoftIslamNPR19,corrabstaxonomy,icse20DLREPAIR}, we use two representative data sources (i.e., \emph{Stack Overflow} and \emph{GitHub}) to study the faults in mobile deep learning. 
Following previous studies~\cite{GuoCXMHLLZL19,CHENDLDEPLOY}, we focus on two representative DL frameworks (i.e., TF Lite and Core ML) that are specially designed for deploying DL models on mobile devices. Since the deployment process is supported by these frameworks, we collect the faults that occur in their usage to construct the dataset of interest.

\subsubsection{Mining SO} 
As one of the most popular community-driven Q\&A websites, SO's users range from novices to experts~\cite{isstaZhangCCXZ18}, increasing the diversity of our collected faults. In addition, developers often post questions on SO for the faults that they cannot find solutions quickly, leading to more non-trivial faults in our dataset.
% To study faults in mobile DL software, we collect relevant questions posted on Stack Overflow (SO), where developers seek technological advice about unresolved issues. 
We collect the relevant questions on SO in the following steps.

\textbf{Download SO dataset.} We first download the entire SO dataset from the official Stack Exchange Data Dump~\cite{solink} on June 7, 2020. The dataset covers the SO posts generated from July 31, 2008 to June 2, 2020. 
% The metadata of each post includes its identifier, post type (i.e., question or answer), creation date, tags, title, body, identifier of the accepted answer if the post is a question, etc. 
Each SO question has one to five tags based on its topics. 
% The developer who posted a question can mark an answer as an accepted answer to indicate that it works for the question.

\textbf{Extract candidate posts.} We then extract SO questions tagged with TF Lite and Core ML. In line with previous work~\cite{isstaZhangCCXZ18,sigsoftIslamNPR19}, we filter out the questions that do not contain any source code because questions about faults usually contain code snippets. In addition, we follow previous studies~\cite{corrabstaxonomy,YILINGBUILD} to exclude questions that do not have an accepted answer, ensuring that we consider only questions with a confirmed solution. As a result, we obtain 154 questions for TF Lite and 149 questions for Core ML.

\subsubsection{Mining GitHub} In addition to SO, GitHub is also a commonly used data source for studying faults. Following previous work~\cite{kbseFrancoGR17}, we mine issues in the official GitHub repositories of the selected frameworks to identify faults that occur during their usage. Compared to commits, issues contain more fault information that includes original reports and developers' discussions~\cite{kbseFrancoGR17}. 
Such a benign characteristic makes issues suitable for studying fault symptoms and fix strategies. In practice, we use the GitHub search API~\cite{githublink} to mine the issues about TF Lite and Core ML on June 27, 2020. 
% The API can take a search query as an input and fetch issues that match the search query from GitHub repositories.
Note that, on GitHub, issues are used for various purposes, including bug report, feature request, etc. To categorize the purposes of issues, developers often employ repository-specific keywords to label issues. In line with previous work~\cite{kbseFrancoGR17}, we also employ the issue labels to help us filter out irrelevant issues. The collection processes for TF Lite and Core ML are conducted separately as follows.

\textbf{Extract issues for TF Lite.} Since TF Lite has been integrated into the TF ecosystem, to obtain issues for TF Lite, we limit the search to issues in the official TF repository~\cite{tfliterepolink}. The first two authors jointly examine each label in the TF repository to determine which labels can be used for filtering. Then, we collect TF Lite related issues by extracting issues labeled with ``\emph{comp:lite}.'' Moreover, we filter out issues labeled with ``\emph{type:feature},'' ``\emph{type:bug},'' ``\emph{type:docs-bug},'' ``\emph{type:docs-feature},'' or ``\emph{type:build/install}'' to exclude those about requests for new features, bugs in the framework itself, document-related problems, and requests for framework installment/build instructions. To ensure that we consider only issues with a confirmed solution, we further exclude those without answers or responses (i.e., those labeled with ``\emph{stalled}'' or ``\emph{stat:awaiting response}''). Overall, we obtain 626 issues for TF Lite.
% Overall, there are a total of 1,463 issues, among which 94 are labeled with ``type:feature,'' 297 with ``type:bug,'' 39 with ``type:docs-bug,'' 83 with ``type:build/install,'' 430 with ``stat:awaiting response,'' and 33 with ``stalled.'' Considering that 106 issues are labeled with more than one label,we finally obtain 626 issues.

\textbf{Extract issues for Core ML.} To obtain issues for Core ML, we first extract all the issues in the official Core ML repository~\cite{coremlrepolink}. Since labels in this repository are not as abundant as those in the TF repository, with the help of issue labels, we can filter out only the issues about bugs in the framework itself (i.e., those labeled with ``\emph{label:bug}''). Then, similar to the process for TF Lite, we extract only the closed issues. Overall, we obtain 169 issues for Core ML.

% \textbf{Extract issues for Core ML.} To obtain issues for Core ML, we limit the search to the issues in official Core ML repository\cite{coremlrepolink} using the argument  "repo:apple/coremltools". Since labels in Core ML repository are not as abundant as tensorflow repository, we collected all closed issues and only filtered out bug reports by using the argument "-label:bug". We collected the data on June 27, 2020 and finally obtained 169 issues for Core ML.

\subsubsection{Refining Dataset} 
Since the extracted posts (i.e., questions and issues) may contain some noise that is not about faults (e.g., how-to questions on SO), the third and fourth authors further filter the extracted posts through manual analysis. Specifically, they jointly read the extracted posts and exclude any post that either is not related to any issue-fixing activity or happens to fix an issue in the framework itself rather than in mobile DL apps. During this process, any conflicts are discussed and resolved by introducing an arbitrator, who has three years of experience in deploying DL models on mobile devices and has published several papers related to this topic in top-tier conferences.
% In line with previous work~\cite{corrabstaxonomy,kbseFrancoGR17}, we also exclude the posts about framework installment/build issues. 
Finally, for TF Lite, we have 65 SO questions and 132 GitHub issues;  for Core ML, we have 52 SO questions and 38 GitHub issues. 
% Considering that some posts have more than one fault, we have a total of 287 posts 304 faults for further analysis.

\subsection{Manual Labelling}
The refined dataset, which consists of 287 posts, is used for distilling symptoms and fix strategies through manual labelling. The scale of this dataset is comparable and even larger than those used in existing fault-related studies~\cite{isstaZhangCCXZ18,issredeep19,iwpcBeyerM0P18,icseAghajaniNVLMBL19,kbseFrancoGR17} that also require manual inspection. Next, we present our procedures of manual labelling.

\subsubsection{Pilot Labelling}
First, we randomly sample 50\% of the 287 posts for a pilot labelling. The first two authors, who have five and three years of DL experience respectively, jointly participate in the process. They follow an open coding procedure~\cite{tseSeaman99} to inductively create categories for symptoms and fix strategies by analyzing the sampled posts. The detailed procedures are described below. 

The two authors read and reread all the posts to understand the context of faults and assign each post with short but descriptive phrases as initial codes to indicate (i) the \emph{fault symptom} that shows what the fault looks like and (ii) the \emph{fix strategy} that tells how a fault is fixed. In this process, they take all the contents of each post, including the title, description, code snippets, error messages, comments, answers, and even URLs mentioned by developers, for careful inspection. 

Then, they proceed to construct taxonomies for symptoms and fix strategies, respectively. Specifically, they group similar codes into categories and the grouping process is iterative, in which they continuously go back and forth between categories and posts to refine the taxonomies. A post is assigned to all related categories if it is related to multiple faults. In the cases where there is no agreement between the two authors, the aforementioned arbitrator is introduced to make discussions and resolve the conflicts. They follow the procedure until they reach agreement on all posts.

\subsubsection{Reliability Analysis} For reliability analysis, the first two authors then independently label the remaining 50\% posts based on the coding schema generated in the pilot labelling. Specifically, they label each post with identified symptom and fix strategy categories and add the posts that cannot be classified into the current taxonomies into a new category named \emph{Pending}. To measure the inter-rater agreement during the independent labelling, we employ the widely used Cohen's Kappa ($\kappa$)~\cite{cohen1960coefficient} as the indicator. The $\kappa$ values obtained for symptoms and fix strategies are 0.819 and 0.743, indicating almost perfect agreement and substantial agreement~\cite{landis1977measurement}, respectively. The agreement levels demonstrate the reliability of our coding schema and procedure. 

The conflicts of labelling are then discussed and resolved by the aforementioned arbitrator. For the posts classified as \emph{Pending}, we also employ the arbitrator to help us further identify symptoms and fix strategies behind them and determine if new categories need to be added. As a result, we add three new categories into the symptom taxonomy and two new categories into the fix strategy taxonomy, and assign all the posts in \emph{Pending} into the taxonomies. 
% Moreover, we check whether the posts that are previously categorized are more fit for the newly-emerging categories. 
The final labelling results are checked and approved by all participants.

In summary, among the 287 posts, we identify a total of 304 faults.
% The entire manual labelling process takes about 400 man-hours. 
The labelling results in pilot labelling and reliability analysis are both included in the final taxonomies. Based on the taxonomies for symptoms and fix strategies, we answer the \textbf{RQ1} and \textbf{RQ2} raised in Section~\ref{back}, respectively.

\section{RQ1: Symptoms}\label{symptom}
\begin{figure*}[h]
\centerline{\includegraphics[width=1.8\columnwidth]{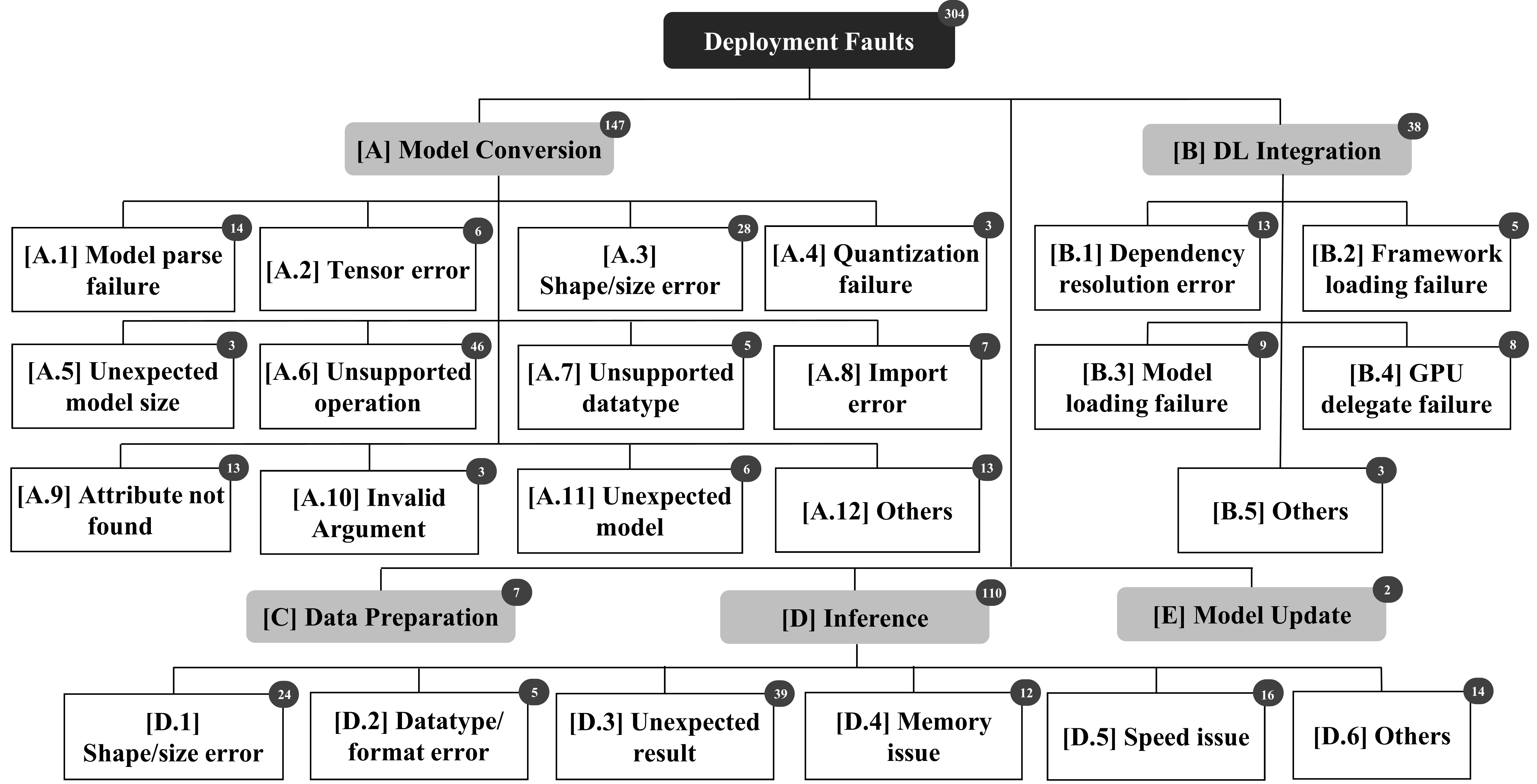}}
\caption{Taxonomy of deployment fault symptoms of mobile DL apps.}\label{fig:tax_symp}
\end{figure*}

Fig.~\ref{fig:tax_symp} presents the hierarchical taxonomy of deployment fault symptoms of mobile DL apps. The taxonomy is organized into three-level categories, including a root category (i.e., \emph{Deployment Faults}), five inner categories linked to stages in deploying DL models (e.g., \emph{Model Conversion}), and 23 specific leaf categories (e.g., \emph{Model parse failure}). 

% The broad spectrum of fault symptoms indicates the prevalence and the diversity of deployment faults in mobile DL software.

\underline{\textbf{Finding 1:}} \emph{We construct a taxonomy of 23 fault symptom categories related to deploying DL models on mobile devices, indicating the diversity of deployment faults.}

% \begin{tcolorbox}
% \textbf{Finding 1:} 
% \end{tcolorbox}

For each category, the number in the top right corner refers to the number of faults in it. Due to space limit, we address only frequent and non-trivial symptoms (i.e., \#faults $\geq$ 3). For \emph{Data Preparation} and \emph{Model Update}, we do not present their leaf categories since no frequent symptoms are observed under them. For the remaining three inner categories, faults with infrequent or unclear symptoms are included in the \emph{Others} category. Next, we discuss and exemplify each inner category.

\subsection{Model Conversion}\label{symcon}
As the first stage of deploying DL models, model conversion aims to convert DL models into the formats expected by mobile devices. 
% Core ML provides specific Python APIs for this task, while TF Lite provides both CLIs and Python APIs. 
To implement a converter for model conversion, developers need to provide the DL model that is ready to be converted and specify necessary information about the model through APIs/CLIs provided by TF Lite or Core ML. We observe 147 faults that occur during the model conversion stage, accounting for 48.4\% of all the identified faults and covering 12 symptom categories.

A large proportion of faults occur when the converter parses the DL model and validates the model information specified by developers, such as names and shapes of input/output tensors of the model. Specifically, 9.5\% of faults in \emph{Model Conversion} are triggered when the converter fails in parsing the DL model (\emph{A.1}).
% often due to incorrect specification of the model path and unresolved model formats. 
% Moreover, the converter also needs to validate the model information specified by developers, such as names and shapes of the input/output tensor of the model. Missing or  incorrect specification  of  these information may result in Tensor  error (A.2) and Shape/size error (A.3).  
Moreover, when the converter detects missing or incorrect specification of the aforementioned model information, developers may encounter \emph{Tensor error (A.2)} and \emph{Shape/size error (A.3)}. Furthermore, \emph{Shape/size error (A.3)} can also be triggered when the converter detects the invalid shape of input/output tensors or the dimension/size misalignment in the model structure. In total, \emph{A.2} and \emph{A.3} account for 23.1\% of faults in \emph{Model Conversion}. In addition to the basic model information, developers can also specify some information to reduce the precision representations of model weights during the conversion stage, so as to reduce the memory cost and computing overhead of DL models on mobile devices. This process is commonly known as model quantization~\cite{CHENDLDEPLOY}. The problematic configuration of quantization-related arguments may result in two types of symptoms, i.e., \emph{Quantization failure (A.4)} and \emph{Unexpected model size (A.5)}, accounting for 6 out of the 147 faults (4.1\%) in \emph{Model Conversion}.

After parsing the DL model, the converter may find that the model uses operations or datatypes that are not supported by TF Lite or Core ML. This can result in \emph{Unsupported operation (A.6)} and \emph{Unsupported datatype (A.7)}, accounting for 31.3\% and 3.4\% of faults in \emph{Model Conversion}, respectively. In particular, \emph{A.6} is the most frequent category in the model conversion stage. Its common occurrence is because that compared to the frameworks used for developing DL models (e.g., TF and Keras), TF Lite and Core ML are proposed later and relatively unfledged. Therefore, some standard operators, functions, or layers (collectively referred to as ``operations'' here) used in the model may be not supported by TF Lite and Core ML. Moreover, the DL model may contain some custom operations that cannot be recognized by the converter. 
% Similarly, some datatypes are also reported to be unsupported.

% The most frequent category under \emph{Model Conversion} is \emph{Unsupported operation (A.6)}, representing 31.3\% of the 147 faults. Compared to the frameworks used for developing DL models (e.g., TF and Keras), TF Lite and Core ML are proposed later and relatively unfledged. Therefore, some standard operators, functions, or layers used in the model may be not supported by TF Lite and Core ML. Moreover, the DL model may contain some custom operations. If not handled properly, these custom operations also cannot be recognized during model conversion. Besides operations, some datatypes are also reported to be unsupported by TF Lite and Core ML, i.e., \emph{Unsupported datatype (A.7)}.

In addition to the symptoms specific to the deployment of DL models, we also observe that a portion (i.e., 15.6\%) of faults in \emph{Model Conversion} share common symptoms with general software systems. For example, 4.8\% of faults are triggered due to unsuccessful import of dependent modules (i.e., \emph{Import error (A.8)}); 8.8\% are related to reference to non-existent variables or functions (i.e., \emph{Attribute not found (A.9)}); and 2.0\% are caused by using arguments of API/CLI incorrectly (i.e., \emph{Invalid argument (A.10)}).
% There are also some faults that are common and similar to those in other computer programs, such as \emph{Import error} (\emph{A.8}, failure in importing specified Python modules), \emph{Attribute not found} (\emph{A.9}, referencing a non-exist class field or function), and \emph{Invalid argument} (\emph{A.10}, argument not satisfying the requirement). These symptoms account for 15.6\% of faults under \emph{Model Conversion}.

Besides the faults with explicit errors thrown during the model conversion stage, sometimes developers get unexpected models even after model conversion appears to be successfully done. For example, developers may find that the number, shape, or format of input/output tensors of the model changes. We classify these cases into the category \emph{Unexpected  model (A.11)}, accounting for 4.1\% faults in \emph{Model Conversion}.
% Sometimes, although no errors occur during the conversion phase, developers still cannot obtain their expected models, i.e., \emph{Unexpected model (A.11)}. Specifically, some developers reported that the number, shape, or format of the input/output tensor changed after conversion although no errors are thrown. Such cases account for 4.1\% faults in \emph{Model Conversion}.

\underline{\textbf{Finding 2:}}
\emph{Most (i.e., 48.4\%) of deployment faults occur during the model conversion stage, covering a wide spectrum of symptoms (i.e., 12 categories). Among these categories, unsupported operation is the most common, accounting for 31.3\% of faults in this stage.}
% \emph{Model conversion is the stage where the most faults occur, accounting for 48.4\% of the total and covering 12 symptom categories. Among these categories, unsupported operation is the most common one, accounting for 31.1\% of faults in this stage.}
% Moreover, tensor error and shape/size error account for a large proportion (23.1\%) of faults that occur in this stage, which mainly relate to improper specification of tensor names and shapes during model conversion.}

\subsection{DL Integration}
After the DL model is converted into the expected format, developers can integrate it as well as DL frameworks into a mobile app project. Then, they can build the project and load the model to make it ready for inference. Faults that appear in this stage are included in the \emph{DL Integration (B)} category, accounting for 12.5\% of the deployment faults of mobile DL apps.

\emph{Dependency resolution error (B.1)} is a common fault when building projects, accounting for 34.2\% of the faults in \emph{DL Integration}. Specifically, it refers to failures in preparing necessary dependencies directly or transitively specified by developers. In these cases, projects throw error messages like inability to resolve libraries, unsuccessful dependency downloading, and undefined reference to objects (e.g., functions and libraries). 

After building projects, developers can run mobile apps to make it predictable. However, in this phase, many developers encounter \emph{Framework loading failure (B.2)} and \emph{Model loading failure (B.3)}, which refer to the failures in loading DL frameworks and models respectively and account for a total of 36.8\% of faults in \emph{DL Integration}.
What is more, developers may configure projects to make it able to use the GPU backend on mobile devices. However, some developers complain that they encounter the \emph{GPU delegate failure (B.4)} when running mobile DL apps. \emph{B.4} represents 21.1\% of faults in \emph{DL Integration}.

\underline{\textbf{Finding 3:}} \emph{Faults appearing in the DL integration stage account for 12.5\% of the total deployment faults and cover five symptom categories. A large proportion (34.2\%) of these faults are thrown with dependency resolution errors.}

\subsection{Data Preparation}
\emph{Data Preparation (C)} is the stage where a mobile app prepares input data for the next inference stage. For a mobile DL app, input data are usually extracted from user-generated data such as camera pictures or typed texts, and a data preparation fault often occurs when the app fails to access or process the required user-generated data. Note that this type of faults is essentially related to data accessing and processing issues, which not only occur in mobile DL apps, but also is very common in other mobile apps.
Therefore, to seek more extensive help, developers usually do not describe these problems in the context of mobile DL apps (e.g., on SO they prefer not to post their problems with any tag related to DL), and thus we observe only a few related cases (2.3\%) with no frequent symptoms in the collected data.
% \emph{Data Preparation (C)} contains the faults that occur when the app extracts input data for the deployed model. In fact, it is a general challenge in mobile app development, not specific to mobile DL apps. For example, developers for a mobile DL app that uses camera pictures as model input may have difficulty in implementing the function of capturing pictures from cameras. This is also a challenge in the development of apps that make use of camera pictures for non-DL purposes. Therefore, to make more developers pay attention to such problems and provide useful suggestions, mobile DL app developers do not need to report such difficulties or related faults in the context of mobile DL apps. As a result, we observe just a few related cases (2.3\%) in the collected data, showing no frequent symptoms.

\subsection{Inference}
\emph{Inference (D)} consists of faults that occur when a mobile app makes inference based on input data. 36.2\% of deployment faults do not show symptoms until this stage. 

A proportion (26.4\%) of faults in \emph{Inference} appear with explicit errors, i.e., \emph{Shape/size error (D.1)} or \emph{Datatype/format error (D.2)}. They are triggered when the shape/size or datatype/format of input/output arrays used for storing input/output data does not align with that of input/output tensors of the DL model.

Furthermore, some developers report that the mobile DL app produces unexpected results (i.e., \emph{D.3}) although no errors are thrown. These cases account for 35.5\% in \emph{Inference}. Specifically, developers may observe that the mobile DL app produces different results than the original model. However, note that this symptom cannot be always used as the indication of faults, especially when model quantization is performed during the model conversion stage. Since model quantization reduces the precision representations of model weights, it is reasonable to observe the change in model performance. Besides, developers also employ some other indications to confirm the existence of \emph{Unexpected result (D.3)}. For instance, the mobile DL app produces the same result for any input or produces different results for the same input.

In addition to the faults that affect the output results, there are also 25.5\% of faults that have impact on the memory usage and inference speed of mobile DL apps. We use \emph{Memory issue (D.4)} and \emph{Speed issue (D.5)} to refer to the two types of faults. Specifically, \emph{Memory issue (D.4)} includes symptoms such as out of memory, memory leak, failures in memory allocation, and segment faults; \emph{Speed issue (D.5)} is mainly manifested as long latency time of making inference.

\underline{\textbf{Finding 4:}} \emph{36.2\% of faults occur when mobile DL apps make inference based on input data, covering six symptom categories. In particular, 35.5\% of the faults in this stage are captured since developers observe unexpected results.}

\subsection{Model Update}
Once put into real usage, mobile DL apps keep receiving feedback from users (e.g., bad cases), based on which DL models can further be improved (e.g., updating the weights of models). 
Instead of re-training DL models on PC/server platforms and then re-deploying the new models again, developers can also directly re-train the DL models on mobile devices, which is the stage \emph{Model Update (E)}. However, since currently on-device training requires a large amount of computational resources and is still not widely supported by existing DL frameworks, we observe only a few instances (0.7\%) related to it in our dataset.
% Once mobile DL apps are really used, they can receive some feedback (e.g., bad cases) that can be used to update weights of the DL models in the mobile apps for further performance improvement. To achieve this, there are two strategies. The first strategy is to re-train the DL models on PC or server platforms. Then developers can deploy the updated models into a new version of mobile app and ship an app store update. The second strategy is to update the models directly via training on mobile devices, which may introduce different faults from other stages. We include such faults in \emph{Model Update (E)}. Since training DL models requires a large amount of computational resources, it is not common to perform it on mobile devices. Moreover, existing frameworks do not support on-device training well. As a result, we observe only a few attempts (0.6\%) at it in our dataset.
% \emph{Model Update (E)} accounts for a trivial proportion (0.6\%) and thus does not show any frequent symptom in our dataset. It consists of the faults that appear when developers directly update the weights of models on mobile devices. Such an update saves developers from shipping an app store update. However, it requires a certain amount of computing power,  memory, and energy of mobile devices. Moreover, existing frameworks do not support it well. As a result, we observe only a few attempts at such an operation in our dataset. 

\subsection{Distribution of Symptoms across Frameworks}
% In this section, we put together the faults collected by considering TF Lite and Core ML to derive the taxonomy. 
We then further analyze the distribution of the identified fault symptoms across the two selected frameworks (i.e., TF Lite and Core ML). We find that the two frameworks share a similar distribution in most categories. For example, 37\% of TF Lite related issues are included in \emph{Inference (D)}, and for Core-ML, the ratio (35\%) is comparable. However, at the same time, the two frameworks differ obviously in some specific categories. For instance, \emph{Unsupported operation (A.6)} accounts for 18\% of TF Lite related issues, but only 10\% of Core ML related issues. 

\section{RQ2: Fix Strategies}\label{fixpattern}
To capture how developers fix different types of deployment faults, for each symptom category, we summarize its fix strategies in this section. Since \emph{Data Preparation} and \emph{Model Update} contain only a few samples and do not show frequent symptoms, here, we do not consider them. 
For the remaining three inner categories, we show the frequency of different fix strategies on their leaf categories in Figs.~\ref{fig:fixpattern1}, \ref{fig:fixpattern2}, and \ref{fig:fixpattern3}, respectively. Due to space limit, strategies with low frequency (i.e., \#faults $\textless$ 3) are not shown in the figures. In each figure, X axis represents each leaf category and the letter identifier is consistent with our taxonomy in Fig.~\ref{fig:tax_symp}; Y axis shows fix strategies following with their total frequency under the inner category. Next, we elaborate the identified fix strategies for frequent symptoms and demonstrate some real-world examples of faults and corresponding fixes.

\subsection{Fix Strategies for Faults in Model Conversion}\label{fixconversion}
\begin{figure}[h]
\centerline{\includegraphics[width=1.0\columnwidth]{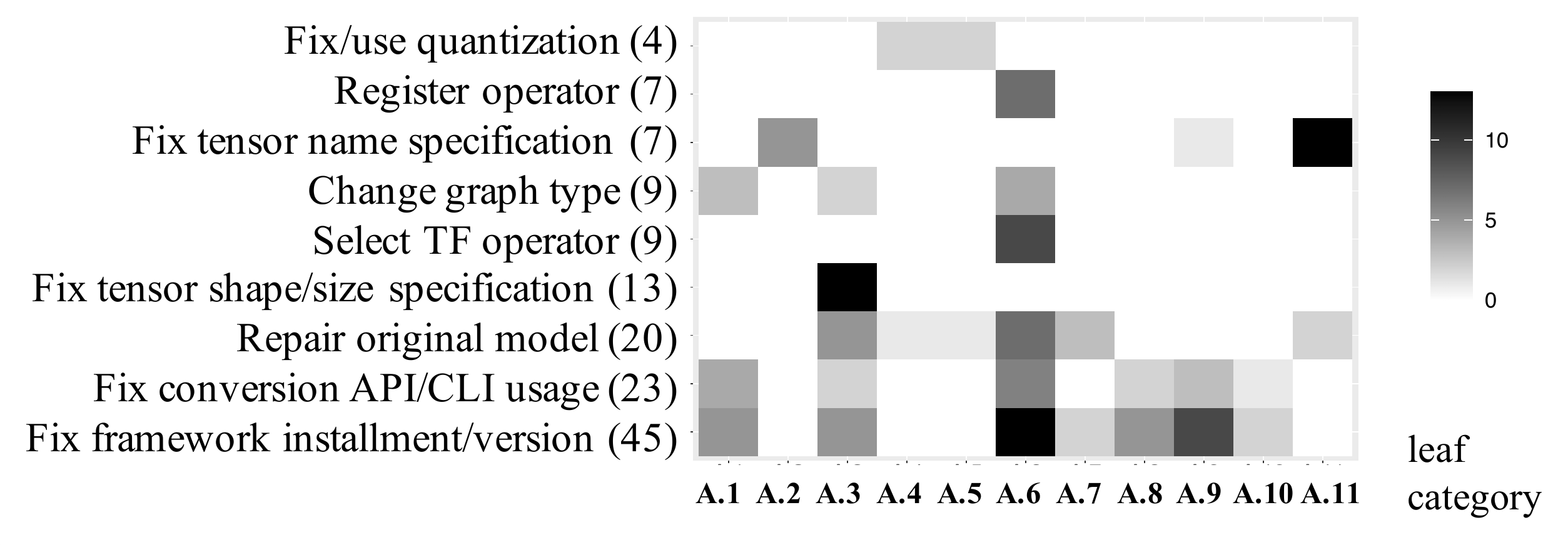}}
\caption{Distribution of fix strategies for leaf categories in \emph{Model Conversion}.}\label{fig:fixpattern1}
\end{figure}

We identify nine frequent fix strategies for faults in \emph{Model Conversion} and illustrate the distribution of these strategies on leaf categories in Fig.~\ref{fig:fixpattern1}. 

\textbf{Fix framework installment/version.} 30.6\% of faults in \emph{Model Conversion} are solved by re-installing the DL framework or switching the DL framework into a different version.
This strategy covers seven fault symptoms, and is especially frequently adopted in the \emph{Unsupported operation (A.6)} and \emph{Attribute not found (A.9)} categories. For example, 36.1\% of \emph{Unsupported operation (A.6)} faults are fixed after switching the DL framework into a more recent version with more supported operations. As for \emph{Attribute not found (A.9)} faults, developers often misuse APIs in a way unsupported by the current DL framework, since APIs frequently evolve with DL frameworks. Therefore, at most cases, developers resolve them by changing the DL framework to another version that supports the reference to specified attributes. For example, a developer reports that she receives an error ``\emph{AttributeError: type object TFLiteConverter has no attribute from\_keras\_model}'' when converting a Keras model to the TF Lite format (TF issue \#38786), and the corresponding fix is upgrading TF to 2.x version since \emph{from\_keras\_model} is not supported by 1.x version.
In addition, the framework version issue can also result in some non-intuitive symptoms. For example, a developer encounters a \emph{Shape/size error (A.3)} during model conversion with the message ``\emph{Check failed: input\_shape.dims().size() == op$\rightarrow$size.size() (4 vs. 3)}'' (SO post \#56631820), which leads to a heated discussion. All the comments suggest that the developer should fix the shape of the input tensor specified during model conversion, but none of them work. Finally, the developer upgrades TF and successfully resolves the fault.

\textbf{Fix conversion API/CLI usage.} 
15.6\% of faults in \emph{Model Conversion}, involving six frequent symptom categories, are fixed by correcting or changing the usage of APIs/CLIs for model conversion. As suggested by previous work~\cite{CHENDLDEPLOY}, so many APIs/CLIs provided by existing DL frameworks for model conversion make it difficult for developers to correctly choose or use their desired APIs/CLIs; meanwhile, frequent addition, deprecation, and upgrade of APIs/CLIs also make their usage error-prone.
% This group of fixes changes the selection or corrects the usage of the API/CLI used for model conversion. 15.6\% of faults under \emph{Model Conversion} can be resolved by it. On one hand, as demonstrated by Chen \textit{et al.}~\cite{CHENDLDEPLOY}, there are so many APIs/CLIs provided by DL frameworks for model conversion, making it difficult for developers to select and use these APIs/CLIs correctly according to their demand. On the other hand, the addition, deprecation, and upgrade of APIs/CLIs caused by framework update  also makes the selection and usage error-prone. Since misuse of APIs/CLIs can result in various faults, this group of fixes covers six frequent symptoms in \emph{Model Conversion}. 

\textbf{Repair original model.} Repairing the DL model used for conversion fixes 13.6\% of faults in \emph{Model Conversion}, which mainly belong to the \emph{Shape/size error (A.3)} and \emph{Unsupported operation (A.6)} categories. As shown in Example (a), the Core ML issue \#525 is a real-world example on the \emph{Shape/size error (A.3)}. A developer uses Keras to implement a binary classifier, trains and tests it successfully. However, when she converts the obtained model to the Core ML format, \emph{Shape/size error (A.3)} occurs. Since she specifies two output labels (``0'' and ``1'') during model conversion, the converter expects a model with a two-dimensional output tensor. However, the output of the original model is an one-dimensional tensor, indicating the probability that the input is classified as label ``1.'' To resolve this fault, the developer repairs the original model and makes it output a two-dimensional tensor, with each dimension indicating the probability that the input is classified as one label (``0'' or ``1'').  As for \emph{Unsupported operation (A.6)}, developers often (i) replace it with a supported one, (ii) implement its function outside the model, or (iii) delete it if it is unnecessary.
% This group of fixes makes some changes to the original DL model used for deployment and solves 13.7\% of faults in \emph{Model Conversion}. More specifically, it mainly resolves the \emph{Shape/size error (A.3)} and \emph{Unsupported operation (A.6)}. For the shape/size error, we take the Core ML issue \#525 as an example. As illustrated in Example (a), a developer used Keras to implement a binary classifier, trained and tested it successfully. However, when she converted the obtained model to the Core ML format, \emph{Shape/size error (A.3)} occurred. Since she specified two output labels (``0'' and ``1'') during model conversion, the converter expected a model with a two-dimensional output tensor. However, the output of the original model was an one-dimensional tensor, indicating the probability that the input is classified as label ``1.'' To resolve such a fault, the developer repaired the original model and made it output a two-dimensional tensor, each dimension indicating the probability that the input is classified as one label (``0'' or ``1''). For the unsupported operation, developers attempt to replace it with a supported one, remove it and implement its function outside the model, or directly remove it from the model if it is not such necessary.
% For the unsupported operation, developers can replace it with a supported one (TF issue \#30872), remove it and implement its function outside the model (SO post \#56221120), or directly remove it from the model if it is not such necessary (SO post \#45943029).

\begin{figure}[t]
    \centering
    \includegraphics[width= 8.7 cm , height= 5.4 cm]{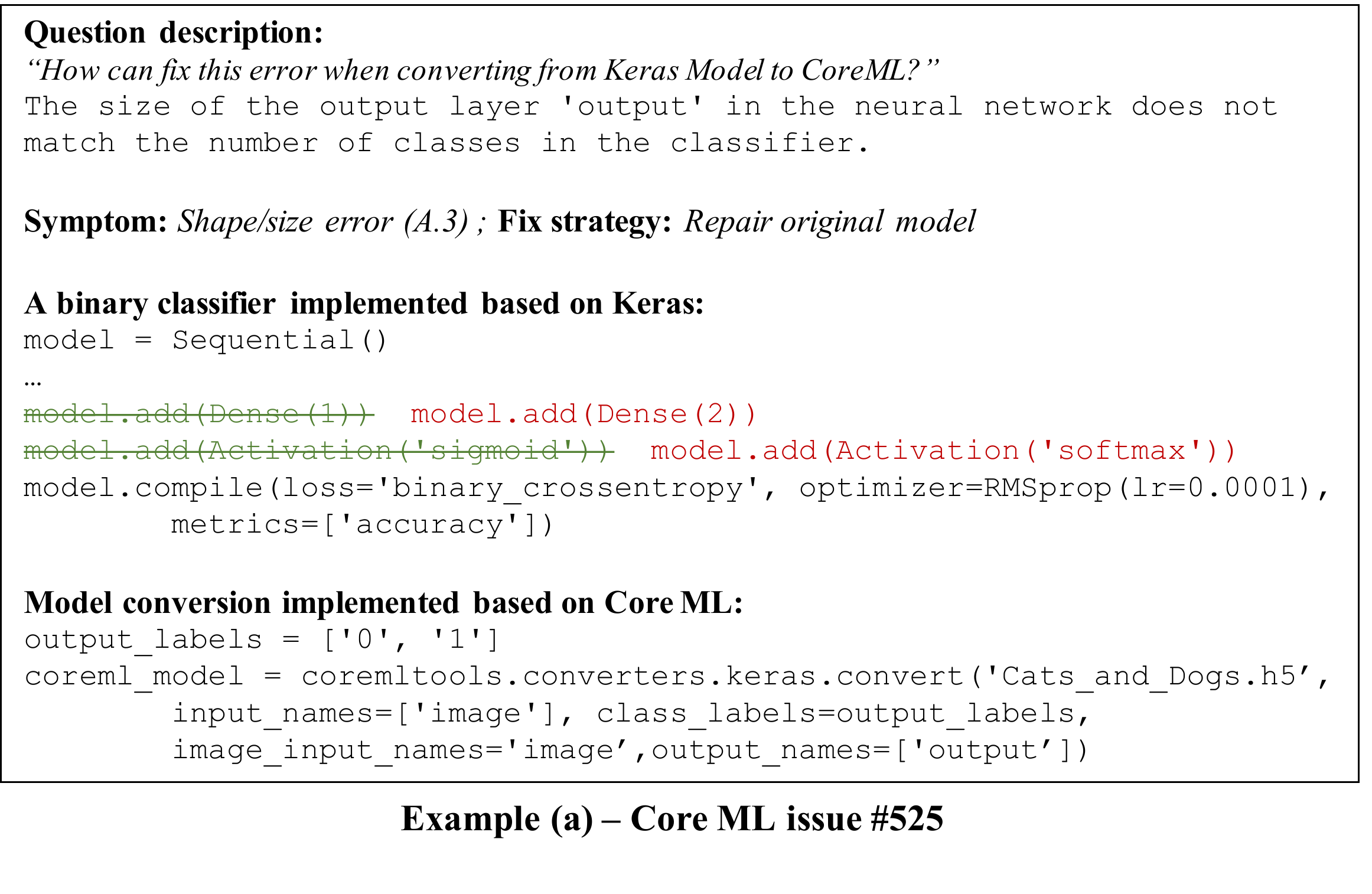}
    \label{fig:ex1}
\end{figure}

\textbf{Fix tensor shape/size specification} \& \textbf{Fix tensor name specification.}
The two strategies fix the specification of the shape/size and the name of input/output tensors during model conversion, respectively. As described in previous work~\cite{icse20DLREPAIR}, training DL models can be expensive since it requires a large quantity of computational resources and labelled data that might not be readily available. Therefore, developers often directly use pre-trained DL models that are available online. In this case, they may have no idea about the model information (e.g., the shape/size and the name of input/output tensors) that needs to be specified during model conversion. Incorrect specification can result in \emph{Shape/size error (A.3)}, \emph{Tensor error (A.2)}, \emph{Unexpected model (A.11)}, etc. Therefore, we can observe that the two strategies mainly fix faults with these symptoms. For example, a developer reuses an object detection model that she is not familiar with from GitHub and specifies the input tensor as a tensor not contained in the model (SO post \#55803971), resulting in \emph{Tensor error (A.2)}. The corresponding solution is fixing tensor name specification.
% The two strategies fix the specification of the shape/size and name of input/output tensors during model conversion, respectively. The former strategy tackles the \emph{Shape/size error (A.3)}, while the latter strategy mainly resolves the \emph{Tensor error (A.2)} and \emph{Unexpected model (A.11)}. As described in previous work~\cite{icse20DLREPAIR}, training DL models can be expensive since it requires a large amount of computational resources and labeled data that might not be readily available. Therefore, developers often use pre-trained DL models that are available online directly. In this case, they may have no idea about the model information that needs to be specified during model conversion. For example, a developer reused an object detection model that she was not familiar with from GitHub and specified the input tensor as a tensor not contained in the model (SO post \#55803971), resulting in \emph{Tensor error (A.2)}. The corresponding solution is fixing tensor name specification.  
% Since incorrect specification of model information is common, the corresponding fix strategies (i.e., fixing the shape/size/name specification) can resolve 13.7\% of faults under \emph{Model Conversion}.

\textbf{Select TF operator} \& \textbf{Register operator.} The two strategies are used to tackle the \emph{Unsupported operation (A.6)} faults that occur when converting DL models into the TF Lite format. Selecting TF operators allows DL models to use a subset of TF operators that are not supported by TF Lite~\cite{tfselect}, while registering operators refers to registering unsupported operators in the TF Lite run-time library so that the run-time knows how to map these operators to executable code~\cite{customop}.
Compared to selecting TF operator, registering operator can be used not only to support TF operators, but also to support operators customized by developers.

\textbf{Change graph type.} This group of fixes changes the type of the model graph (e.g., training graph and evaluation graph) used for conversion. The model graph refers to the computational graph that represents the structure of the DL model. Since operations involved in model training and evaluation are not always the same, developers need to construct the training graph and the evaluation graph separately. The graph used for conversion should be the evaluation graph since developers always aim to make inference rather than training on mobile devices. When developers use the training graph for conversion, some training operations may be unrecognized and unsupported by the converter. As a result, developers would encounter \emph{Unsupported operation (A.6)}. \emph{Model parse failure (A.1)} is another common symptom that occurs when the incorrect type of model graph is provided.
%(SO post \#56707573)
%(SO post \#56779949)

\textbf{Fix/use quantization.} This group of fixes selects a proper quantization method according to developers' demand or fixes the incorrect quantization configuration. Naturally, it can resolve the \emph{Quantization failure (A.4)}. In addition, since model quantization can reduce the model size while reducing the precision representations of model weights, when developers observe that the model size does not change as expected after quantization (i.e., \emph{Unexpected model size (A.5)}), there may be a fault in the quantization configuration that needs to be fixed. 
% For example, a developer complained that as she quantized the model weight from 32-bits floating points to 8-bits integers, theoretically the model size should be reduced by four times (SO post \#57631313). However, the size of the quantized model is the same as that before model quantization. Finally, she fixed the problematic quantization configuration, thereby obtaining a converted model with the expected size. 

\underline{\textbf{Finding 5:}} \emph{We identify nine frequent fix strategies for faults in model conversion. The three most common strategies are fixing framework installment/version, fixing conversion API/CLI usage, and repairing the original model, resolving 30.6\%, 15.6\%, and 13.6\% of faults in this stage, respectively. }

\subsection{Fix Strategies for Faults in DL Integration}\label{fixinte}
\begin{figure}[h]
\centerline{\includegraphics[width=0.9\columnwidth]{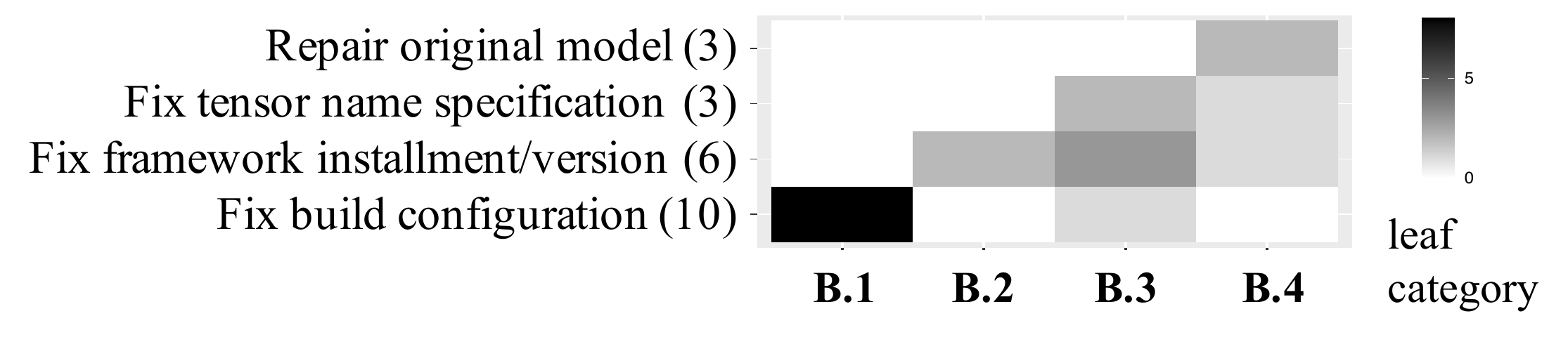}}
\caption{Distribution of fix strategies for leaf categories in \emph{DL Integration}.}\label{fig:fixpattern2}
\end{figure}

As illustrated in Fig.~\ref{fig:fixpattern2}, we identify four frequent fix strategies for faults in \emph{DL Integration}.
% We identify 15 fix patterns for the 38 faults in \emph{DL Integration}, but only four strategies are considered as frequent and illustrated in Fig.~\ref{fig:fixpattern2}. 
% This indicates that many faults in DL Integration are fixed by case-by-case strategies.

\textbf{Fix build configuration.} 26.3\% of faults in \emph{DL Integration} are resolved by fixing the build configuration of mobile DL projects, including fixing dependency version, fixing link configuration, fixing option settings, etc. This fix strategy mainly resolves the \emph{Dependency resolution error (B.1)}.
% It resolves 26.3\% of faults in \emph{DL Integration} and mainly tackles the \emph{Dependency resolution error (A.1)}.

The remaining three frequent fix strategies have been described in Section~\ref{fixconversion}. They are also applicable to some faults in \emph{DL Integration}.

\textbf{Fix framework installment/version.} When the required DL framework is not successfully installed or the DL model is not incompatible with the framework version used in the project, symptoms like \emph{Framework loading failure (B.2)} and \emph{Model loading failure (B.3)} may occur. In such cases, developers need to fix framework installment/version.

\textbf{Fix tensor name specification.} When input/output tensors are specified incorrectly during model conversion, the converted model may not be loaded in mobile projects successfully (i.e., \emph{Model loading failure (B.3)}). Moreover, improper specification of input/output tensors may cause \emph{GPU delegate failure (B.4)}. For instance, a developer encounters this failure since some data pre- and post-processing operators in the original model are not supported by GPU (TF issue \#25238). The fixing strategy is re-specifying the input and output tensors during model conversion to ensure that the unsupported operators are not between the new input and output nodes, thereby not in the converted model.

\textbf{Repair original model.} This strategy can resolve the \emph{GPU delegate failure (B.4)}. In fact, some operators supported by DL frameworks are not supported by GPU. In this case, developers can repair the original model to remove these operators and implement alternative operations, so that the integrated DL models can run on the GPU backend of mobile devices.

\underline{\textbf{Finding 6:}} \emph{We identify four frequent fix strategies for faults in DL integration. The most common one is fixing build configuration, which resolves 26.3\% of faults in this stage.}

\subsection{Fix Strategies for Faults in Inference}
\begin{figure}[h]
\centerline{\includegraphics[width=1.0\columnwidth]{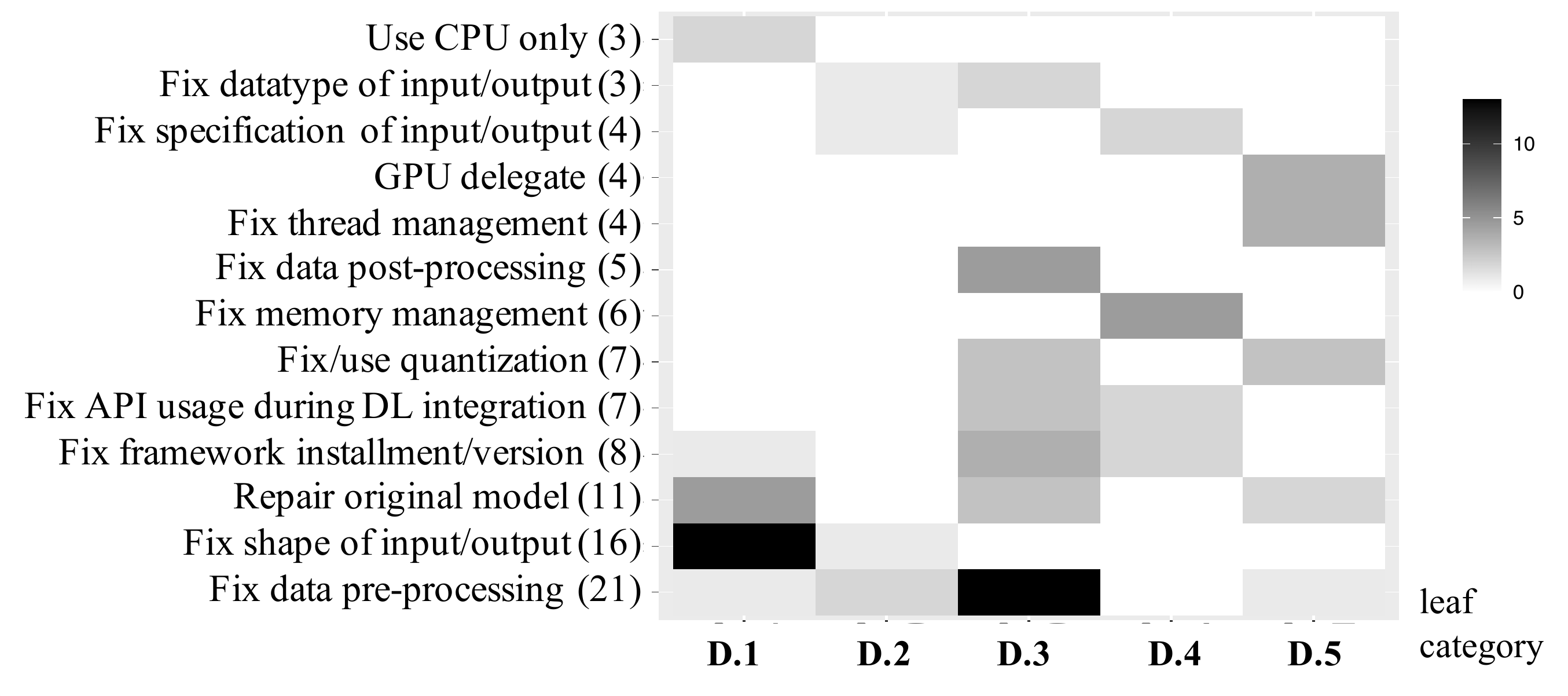}}
\caption{Distribution of fix strategies for leaf categories in \emph{Inference}.}\label{fig:fixpattern3}
\end{figure}

We identify 13 frequent fix strategies for faults in \emph{Inference} and present the distribution of these strategies in Fig.~\ref{fig:fixpattern3}.

\textbf{Fix data pre-processing} \& \textbf{Fix data post-processing.} 23.6\% of faults in \emph{Inference} can be resolved by fixing the process of preparing data for model input (i.e., data pre-processing) or the process of parsing model output to obtain expected or human-readable results (i.e., data post-processing). When developing DL models, data pre-processing is often considered as an individual stage~\cite{AlshangitiSMLY19} and thus may not be included inside the model structure. In this case, code for data pre-processing needs to be re-implemented in the mobile project during the deployment process, so as to keep the consistent behaviors of the DL model before and after deployment. Forgetting to implement it or implementing it incorrectly can result in unexpected results. In addition, sometimes the model behaves well and generates the expected output, but developers make mistakes in parsing the model output, which can also result in unexpected results. Therefore, we can find that the two fix strategies mainly tackle 
the \emph{Unexpected result (D.3)} and 48.7\% of faults in this category can be resolved by them.
% Forgetting to implement it (SO post \#60382477) or implementing it incorrectly (SO post \#56762749) can result in unexpected results. In addition, sometimes the model behaves well and generates the expected output, but developers make mistakes in parsing the model output to get the final result (SO post \#49316297). In this case, they may also obtain the unexpected results. Therefore, we can find that the two fix patterns mainly target at the \emph{Unexpected result (D.3)} and 48.7\% of faults in \emph{Unexpected result (D.3)} can be resolved by them.

\begin{figure}[t]
    \centering
    \includegraphics[width= 7.7cm , height= 7 cm]{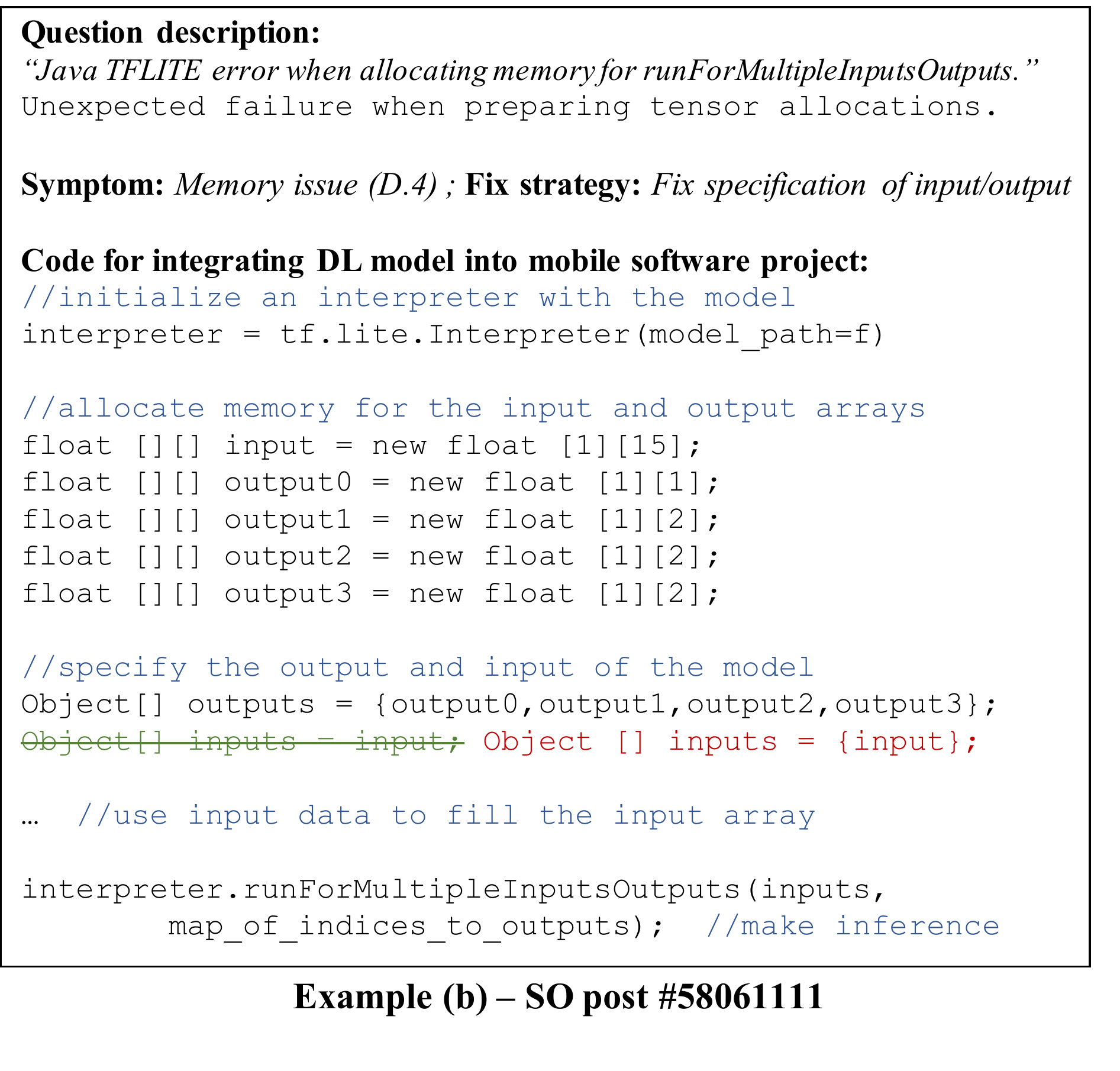}
    \label{fig:ex2}
\end{figure}

\textbf{Fix shape of input/output} \& \textbf{Fix datatype of input/output.} \& \textbf{Fix specification of input/output.} When integrating DL models into mobile projects, developers often need to prepare the input/output arrays that are used for storing input/output data and specify their shape and datatype. For example, as shown in Example (b) (SO post \#58061111), a developer integrates a DL model with one input tensor and four output tensors into an Android project implemented in Java. First, she uses the model to initialize an interpreter. Then, she allocates memory and specifies the shape and datatype for input and output arrays, and sets these arrays as the model input and output. Finally, she uses input data to fill the input array and make inference. During the above process, when the shape of input/output arrays are incorrectly specified, developers may encounter \emph{Shape/size error (D.1)}. Similarly, when the specification of the datatype of input/output arrays is incorrect, developers may encounter \emph{Datatype/format error (D.2)} or obtain \emph{Unexpected result (D.3)}. Therefore, fixing shape of input/output mainly resolves faults in \emph{D.1}, while fixing datatype of input/output tackles faults in \emph{D.2} and \emph{D.3}. In addition, incorrect specification of input/output of the model may result in faults such as \emph{Datatype/format error (D.2)} and \emph{Memory issue (D.4)}. For example, the symptom of the fault in Example (b) is \emph{Memory issue (D.4)} with an error ``\emph{Unexpected failure when preparing tensor allocations}''. The corresponding solution is fixing specification of input/output.

\textbf{Fix API usage during DL integration.} In the DL integration process as shown in Example (b), developers often misuse relevant APIs provided by DL frameworks. The corresponding solution is fixing API usage during DL integration.

\textbf{Fix memory management.} 
This group of fixes resolves the faults related to memory management during the DL integration stage. A typical fault is that developers may set the input/output arrays before allocating memory for them (e.g., SO post \#56819142), which results in \emph{Memory issue (D.4)}. 

\textbf{Fix thread management} \& \textbf{GPU delegate.} The two groups of fixes refer to setting an appropriate number of threads in mobile projects and configuring mobile projects to enable DL models in them to run on the GPU backend, respectively. Both of them can reduce the latency during inference and thus resolve 50\% of the faults in \emph{Speed issue (D.5)}. 

\textbf{Use CPU only.} This group of fixes forces DL models to run on the CPU backend during inference by configuring some settings in mobile projects. It mainly resolves the \emph{Shape/size error (D.1)}. For example, when a developer makes inference with a Core ML model, an error is thrown, reporting that the size of the input sequence exceeds the upper bound (SO post \#52144540). The cause is that a dense operation in the model with a large size of sequences is unable to be performed on a GPU due to the memory constrains. Finally, the developer forces the model to use only CPU and resolves the fault.

In addition, there are three fix strategies that have been elaborated in Sections~\ref{fixconversion} and \ref{fixinte}.

\textbf{Repair original model.} This strategy mainly resolves the \emph{Shape/size error (D.1)} and \emph{Unexpected result (D.3)}. Specifically, when the input size expected by DL model is inconsistent with the actual size of data extracted in apps (i.e., \emph{Shape/size error (D.1)}), one solution is to reshape the original models. Moreover, some developers find that the models cannot perform well in real applications (i.e., \emph{Unexpected result (D.3)}) and thus choose to refine their original models.

\textbf{Fix framework installment/version.} Fixing framework installment/version can also resolve some faults that occur during inference. For example, a developer gets worse results when she converts a Keras model into the TF Lite format (SO post \#51966486). The root cause is that an API that she uses during model conversion is problematic in TF 1.10. Upgrading TF to version 1.11 resolves the fault.

\textbf{Fix/use quantization.} The problematic configuration of model quantization can affect performance of the converted model and thus result in unexpected inference results. Moreover, since the quantized model is more light than the original one, model quantization is also a solution to speed up the inference process. Therefore, fixing/using quantization mainly resolves the \emph{Unexpected result (D.3)} and \emph{Speed issue (D.5)}.

\underline{\textbf{Finding 7:}} \emph{The fix strategies for faults in inference are diverse. They cover many stages of the deployment process, including fixing data processing, fixing the model conversion stage (e.g., fixing/using quantization), fixing the DL integration stage (e.g., fixing API usage during DL integration), etc.}
\section {Discussion}\label{implication}
Given the rapidly increasing popularity of mobile DL apps, our study has timely and immediate implications for developers, especially novice developers. Specifically, our results can help developers more efficiently understand and resolve common deployment faults.
% aid them in avoiding common pitfalls and addressing common faults that they encounter. 
For example, a developer may be confused as to how to resolve the \emph{Unsupported Operation (A.6)} symptom, since the fault may lie in the model development process or any setting/configuration in model conversion. However, with our results, the developer can know how such faults are usually resolved in practice so that she can find the solution with less trial and error.

Nevertheless, due to the broad spectrum of deployment faults, it is challenging for developers to detect and fix these faults completely manually. Therefore, we call on SE researchers to develop automated techniques to assist them. Although the combinations of fault symptoms and fix strategies derived in our study can serve as common strategies for the automated techniques, we believe that more research efforts are needed to achieve the goal. Next, we discuss some implications of our findings on future research.

\textbf{Testing DL models deployed on mobile devices.}
As suggested in our study, 20.1\% of deployment faults (e.g., \emph{Unexpected model (A.11)}, \emph{Unexpected result (D.3)}, and \emph{Speed issue (D.5)}) do not explicitly lead to an error or a crash during deployment, and are thus usually exposed relying on developers' experience or extra efforts. This non-trivial portion indicates the importance of testing deployed models automatically. However, existing testing efforts~\cite{wcreMaJXLLLZ19,kbseMaJZSXLCSLLZW18,sospPeiCYJ17} are mainly dedicated to the DL models obtained by training, rather than the DL models converted and deployed on mobile devices. Unlike testing the trained DL models, testing deployed DL models on mobile devices has its unique challenges in (i) \emph{resource limitation} and (ii) \emph{undetermined change in model behaviors}. Specifically, compared to the PC/server platforms used for testing trained DL models, mobile devices used for testing deployed models have limited resources in terms of computing power and memory size. In addition, in the cases where quantization techniques are employed during model conversion, the deployed models should have different behaviors from the original models since quantization techniques reduce the precision representations of model weights. However, it is unclear how differently the models after deployment might behave, increasing the difficulty in testing the deployed models. For example, a developer gets worse predictions using a TF Lite model converted from a Keras model (SO post \#51966486). Since she employs quantization techniques during model conversion, it is difficult for her to tell whether the performance loss of the model is caused by only the quantization or other bugs in the deployment process.
To the best of our knowledge, there is little work focusing on the deployed model testing. 
With increasing growth of mobile DL apps, we encourage researchers to conduct research in this direction and propose some testing techniques accordingly.

\textbf{Repairing DL models based on deployment faults.} We can find that repairing the original DL models used for deployment is a common fix strategy for faults that occur in model conversion, DL integration, and inference stages. Specifically, it resolves 11.2\% of deployment faults, covering 10 frequent symptoms. Therefore, we believe that this significant fix strategy deserves the attention of researchers. However, existing research efforts~\cite{icse20DLREPAIR} focus on repairing DL models in the development process and investigate the correlation between different model repairing patterns and various fault types in the development process, including API faults, data faults, structural faults, etc. By comparison, there is little work on repairing DL models based on faults identified in the deployment process. We call on researchers to develop automated techniques in this direction to facilitate the automated fix of deployment faults of mobile DL apps.

\textbf{Mining API/CLI usage protocols.} In this study, we observe that 34 out of 304 faults are resolved by fixing API/CLI usage in model conversion and DL integration stages. Mining the API/CLI usage protocols enforced by DL frameworks is a promising research topic to facilitate the automated detection and fixing of these faults. Specifically, researchers can mine these protocols from the official documentation of DL frameworks and relevant projects available on open-source code repositories. In particular, the changes in the API/CLI usage protocols caused by the evolution of DL frameworks need to be highlighted in the mining results.

\section{Threats to Validity}\label{threats}
In this section, we discuss threats to the validity of our study.

\textbf{Selection of frameworks.} Our identification of deployment faults of mobile DL apps is based on two relevant frameworks, which may lead to possible selection bias in this study. To mitigate this threat, we select the representative and widely-used frameworks. On the one hand, the selected frameworks are widely used in industry practice and well adopted in related studies~\cite{CHENDLDEPLOY,GuoCXMHLLZL19}. On the other hand, the selected frameworks cover the deployment scenarios of two typical types of mobile apps (i.e., Android and iOS apps). 

\textbf{Selection of data sources.} Since there is no list of all mobile DL app projects in the world, our study cannot cover all the relevant faults, which may lead to a threat to the external validity. To mitigate this threat, we select two representative data sources (i.e., SO and GitHub) that have been widely used in empirical studies in SE~\cite{isstaZhangCCXZ18,sigsoftIslamNPR19,corrabstaxonomy,icse20DLREPAIR,icseAghajaniNVLMBL19}. Since previous studies~\cite{corrabstaxonomy,icse20document} have found that findings derived from SO and GitHub posts can be well validated by practitioners, we believe that our choice of SO and GitHub does not invalidate our results. However, it is still possible that in other contexts developers may encounter faults that are not covered in this study. In the future, we plan to include interviews with researchers and practitioners to further validate our findings.

\textbf{Subjectivity of researchers.} The subjectivity of researchers presents a possible threat to the validity of manual analysis. To mitigate this threat, we ensure that each case is labelled by at least two authors with an experienced arbitrator resolving the conflicts and inspecting all final results. In addition, the inter-rater agreement is relatively high, which demonstrates the reliability of the labelling schema and procedure.

% \texbf{Timeliness of results.}
\section{Related Work}\label{related}
In this section, we summarize the related work to well position our study within the literature.

\textbf{Challenges that ML/DL poses for SE.} Machine learning (ML) plays an increasingly significant role in various application domains and poses new challenges for software developers~\cite{jiezhangml}. To understand these challenges, Alshangiti et al.~\cite{AlshangitiSMLY19} analyzed the ML-related questions posted on SO and found that these questions are more difficult to answer than other questions. By further analysis, they demonstrated that model deployment is the most challenging across all the ML phases and that DL-related topics are the most common in the ML-related questions. In recent years, several studies have focused on the challenges in developing DL applications. For example, Han et al.~\cite{eseHanSWDX20} applied an automatic topic modelling technique to the SO questions related to three popular DL frameworks and derived the topics contained in these questions. Their results revealed common concerns that developers face when using DL frameworks, such as version problems and model training. Similarly, Zhang et al.~\cite{issredeep19} manually analyzed DL-related questions on SO and found that program crashes, model deployment, and implementation related questions are the most frequently asked. Recently, Chen et al.~\cite{CHENDLDEPLOY} investigated SO questions related to the  deployment process of DL based applications. They derived the topics of the specific challenges that developers face when deploying DL models to server, mobile, and browser platforms. In contrast, instead of deriving the topics of challenges at a macro level, we aim to analyze  symptoms and fix strategies of the deployment faults and provide actionable implications for fault detection and fix in mobile DL apps. In addition, we do not limit our analysis to just SO and also consider GitHub, which ensures comprehensiveness of this study.

\textbf{Empirical study on faults.} There have been a number of empirical studies that focus on faults in different types of software systems. For example, Lu et al.~\cite{asplosLuPSZ08} studied concurrency fault characteristics; Franco et al.~\cite{kbseFrancoGR17} explored real-world faults in numerical software; Gao et al.~\cite{GaoDQGW0HZW18} conducted an empirical study on recovery faults in large-scale distributed systems.
In recent years, the rapid development of DL technologies has inspired some empirical studies on characterizing the faults in software applications that make use of DL frameworks. For example, Zhang et al.~\cite{isstaZhangCCXZ18} collected faults in TF programs from SO and GitHub. They categorized the symptoms and root causes of these faults through manual analysis. Following this work, Humbatova et al.~\cite{corrabstaxonomy} and Islam et al.~\cite{sigsoftIslamNPR19} extended their scope to the faults in programs written based on five popular DL frameworks to present more comprehensive results. Moreover, Islam et al.~\cite{icse20DLREPAIR} analyzed the fix strategies of these faults in their follow-up work. Recently, Zhang et al.~\cite{icse20DLjobs} studied the program faults of DL jobs running on a remote and shared server platform. Across the existing empirical studies, faults are often characterized based on multiple dimensions, including types, symptoms, root causes, fix strategies, etc. Compared to the prior studies, we apply these fault characterization methods to the faults in a different domain, i.e, mobile DL apps.

\textbf{Mobile DL apps.} To make DL models accessible for users, developers need to deploy them to different platforms according to various application scenarios. A popular way is to deploy them on mobile devices. To facilitate this  deployment process, researchers have proposed many optimization techniques (e.g., cloud offloading~\cite{mobicomXuZLLL18} and model compression~\cite{mobisysLiuLZNLD18}). In addition, researchers have built numerous DL based applications on mobile devices~\cite{imwutXuQMHL18,hucRaduLBMMK16,hucMittalYGK16}. To bridge the knowledge gap between research and practice, Xu et al.~\cite{wwwXuLLLLL19} conducted an empirical study on large-scale Android apps collected from Google Play store and demonstrated the increasing popularity of DL in real-world mobile apps. Despite this popularity, the related techniques for deploying DL models to mobile devices are still not very mature. Recently, Guo et al.~\cite{GuoCXMHLLZL19} investigated the performance gap when the trained DL models are migrated from PC to mobile devices with the help of TF Lite and Core ML. Their findings unveiled that the deployment still suffers from compatibility and reliability issues. Despite these efforts, the characteristics of deployment faults of mobile DL apps are still under-investigated and thus we aim to fill in this knowledge gap.
\section{Conclusion}\label{conclusion}
In this paper, we have presented a comprehensive study of deployment faults of mobile DL apps. By manual examination of 304 real-world faults extracted from SO and GitHub, we have derived a taxonomy of fault symptoms with 23 categories, indicating that the process of deploying DL models on mobile devices stretches over a wide spectrum of faults. Moreover, we have analyzed  the fixes for the extracted faults and distilled frequent combinations of fault symptoms and fix strategies that can be adopted to facilitate manual and automated fault fix. Finally, we have discussed insightful implications for developers and researchers based on our results.

\section*{Acknowledgment}
This work was supported by the National Key Research and Development Program of China under the grant number 2018YFB1004403, the National Natural Science Foundation of China under the grant number 61725201, and the Beijing Outstanding Young Scientist Program under the grant number BJJWZYJH01201910001004. Haoyu Wang’s work was supported by the National Natural Science Foundation of China under grant numbers 62072046 and 61702045.

\balance
\bibliographystyle{IEEEtran}
\bibliography{dldeploybib}

\end{document}